# MODELLING ANIMAL-VEHICLE COLLISION COUNTS ACROSS LARGE NETWORKS USING A BAYESIAN HIERARCHICAL MODEL WITH TIME-VARYING PARAMETERS

Krishna Murthy Gurumurthy, Ph.D.
Department of Civil, Architectural and Environmental Engineering
The University of Texas at Austin
gkmurthy10@utexas.edu

Zili Li, Ph.D.
Post-Doctoral Researcher
University of Queensland
Brisbane, Australia
zili.li@uq.edu.au

Kara M. Kockelman, Ph.D., P.E.
Professor and Dewitt Greer Centennial Professor of Transportation Engineering
Department of Civil, Architectural and Environmental Engineering
The University of Texas at Austin – 6.9 E. Cockrell Jr. Hall
Austin, TX 78712-1076
kkockelm@mail.utexas.edu

Prateek Bansal, Ph.D.
Leverhulme Early Career Post-Doctoral Fellow
Imperial College of London
p.bansal19@imperial.ac.uk

**ABSTRACT**

Animal-vehicle collisions (AVCs) are common around the world and result in considerable loss of animal and human life, as well as significant property damage and regular insurance claims. Understanding their occurrence in relation to various contributing factors and being able to identify locations of high risk are valuable to AVC prevention, yielding economic, social and environmental cost savings. However, many challenges exist in the study of AVC datasets. These include seasonality of animal activity, unknown exposure (i.e., the number of animal crossings), very low AVC counts across most sections of extensive roadway networks, and computational burdens that come with discrete response analysis using large datasets. To overcome these challenges, a Bayesian hierarchical model is proposed where the exposure is modeled with nonparametric Dirichlet process, and the number of segment-level AVCs is assumed to follow a Binomial distribution. A Pólya-Gamma augmented Gibbs sampler is derived to estimate the proposed model. By using the AVC data of multiple years across about 100,000 segments of state-controlled highways in Texas, U.S., it is demonstrated that the model is scalable to large datasets, with a preponderance of zeros and clear monthly seasonality in counts, while identifying high-risk locations (for application of design treatments, like separated animal crossings with fencing) and key explanatory factors based on segment-specific factors (such as changes in speed limit) can be done within the modelling framework, which provide useful information for policy-making purposes.

**INTRODUCTION**

Animal-vehicle collisions (AVCs) are common around the world and result in considerable loss of animal and human life, as well as significant property damage and regular insurance claims (Al-Ghamdi and AlGadhi, 2004; Bruinderink and Hazebroek, 2003; Klöcker et al., 2006; Mountrakis and Gunson, 2009; Mrtka and Borkovcová, 2013; Seiler, 2005; Sullivan, 2011). For such reasons, there is continued research in AVC prediction and the effectiveness of various prevention measures (Gunson et al., 2011).

Special attention has been paid to AVCs' spatial and temporal attributes, due to clustering at certain times of year and times of day, with different species' movements and breeding seasons (see, e.g., Wilkins et al. 2019). In the spatial dimension, the focus is on identifying the relationship between AVC locations and animal habitats (Dettki et al., 2011; Gkritza et al., 2010; Hurley et al., 2009) and nearby landscapes (Danks and Porter, 2010; Grilo et al., 2009; Jensen et al., 2014; MALO et al., 2004). In the temporal dimension, within-day and seasonal activity patterns of both animals and humans vary, affecting vehicle presence and animal presence - and their collisions - on roadways (Dettki et al., 2011; Diaz-Varela et al., 2011; Haikonen and Summala, 2001; Rowden et al., 2008). Across the year, migratory patterns, variations in sunrise and sundown, and climatic conditions also play a role (Garrett and Conway, 1999; Hothorn et al., 2015; Niemi et al., 2017; Rodríguez-Morales et al., 2013).

In terms of AVC prevention, the effectiveness of warning signs (Ujvari et al., 2007), light-reflecting devices (Brieger et al., 2016), fencing and barriers (LEBLOND et al., 2007; Zuberogoitia et al., 2015), overpasses and underpasses (McCollister and van Manen, 2010; Rodriguez et al., 2010), modification of nearby landscapes (Jaeger et al., 2016), overhead lighting, and other treatments have been investigated. Some studies have emphasized the effects of roadway design details on AVCs, like speed limit choices (Found and Boyce, 2011; Meisingset et al., 2014), road widths (Litvaitis and Tash, 2008), shoulder widths (Lao et al., 2011a), and the number of lanes used (Lao et al., 2011b). Most of these works have been studied at an aggregate level (zonal or corridor level) to avoid discrete counts and allow researchers to focus on general trends.

A handful of recent studies also have explored the factors affecting the severity levels of AVCs using discrete choice models (Ahmed et al., 2021; Al-Bdairi et al., 2020), but the results of such models are difficult to apply in developing preventive measures because accident-level factors generally dominate operational and planning factors in determining the severity of an AVC. Instead, hotspots are central to assessing AVC clustering on segments in large networks. Two approaches are normally adopted to deal with thousands of distinct locations and network links. A computationally simple approach used by Kolowski and Nielsen (2008) relies on correlation coefficients to define the similarity between road segments with AVC occurrences, and judges hotspots according to correlation strengths. Alternatively, kernel-based smoothing can be applied across all segments at once (Bíl et al., 2016; Ramp et al., 2006; Snow et al., 2014). As noted by Snow et al. (2014), such hotspot identification methods normally require a large number of subjective inputs (like the spatial weights and kernel band-width used) in the implementation process, and can result in unreliable inference. More importantly, the relative importance of each attribute for identifying hotspots is generally unknown, and scenario evaluations based on specific attributes can be unreliable or impossible.

This study improves upon such methods and demonstrates how local hotspots can be identified using a Bayesian binomial regression model on a large-scale network of over 100 thousand segments. A Gibbs sampler is derived to estimate the proposed model. The model facilitates scenario evaluations based on any segment-specific attribute (like speed limit), which could be valuable in developing AVC prevention practices. There are three main challenges in modeling AVC counts. First, traditional count data models cannot be directly used because exposure (i.e., the number of animal road crossings) is unknown. Second, a high proportion of segments have zero AVCs. Third, along with unobserved heterogeneity in the effect of covariates across segments, seasonality and spatial correlation are required to be modeled to capture heterogeneity in AVC counts. No existing discrete response model can address all these challenges simultaneously due to trade-off between computational tractability and flexibility, but several studies could handle them individually. The literature related to each of these modeling challenges and the adopted approach is discussed below.

Whereas traffic volume and its proxies are used as the exposure in crash count data models, exposure is not required in ordered or multinomial response (i.e., injury severity) models. To the best of our knowledge, entirely unknown exposure has not been modeled in in crash count data models. However, Crépet and Tressou (2011) demonstrate how a nonparametric Dirichlet process (DP) mixture can be used to model exposure in food risk analysis. DP mixture has also been used in accident analysis, but to model the semi-parametric heterogeneity in multivariate and multilevel count data models (Heydari et al., 2016, 2017). We illustrate the first application of the DP mixture to model exposure in crash count data models.

High proportion of zeros in discrete responses are generally handled using zero-inflated models (Anastasopoulos, 2016; Fountas and Anastasopoulos, 2018; Liu et al., 2018). However, we do not adjust for preponderance of segments with zero crashes because DP mixture inherently handles such situations. Specifically, DP mixture creates clusters of segments in a data-driven manner and segments in same cluster can share the information about the number of animal crossings.

Extant literature has emerged in the last decade on modeling unobserved heterogeneity in discrete response regression models, such as spatiotemporal correlation in intercept term (Liu and Sharma, 2017, 2018), mixture-of-normal-distributed random parameters to represent cross-sectional unobserved heterogeneity (Xiong and Mannering, 2013; Buddhavarapu et al., 2016; Mannering et al., 2016; Huang et al., 2019), and heterogeneity in mean and variance of mixing distributions (Yu et al., 2019, 2020; Fanyu et al., 2021; Li et al., 2021; Yan et al., 2021). Hou et al. (2021), Krueger et al. (2020a), and Mannering et al. (2020) have argued and illustrated that accounting for unobserved heterogeneity improves the predictive ability of discrete response models. However, Krueger et al. (2020b) also show that such gains in predictive accuracy are compensated when a linear link function is replaced by a nonparametric counterpart in spatial count data models. Considering that the predictive ability of the crash count model is crucial and nonparametric link function would make the estimation time prohibitably large, accounting for unobserved heterogeneity through random parameters in linear link function is in the interest of this study. However, with the large-scale dataset at hand, DP mixture would lead to challenges in the mixing and convergence of the Gibbs sampler (Hastie et al., 2015). Therefore, instead of making link function parameters random, a segment-time-specific random intercept is included in the model (see details in Eq. 3 of "The Modeling Framework" section). Spatial correlation is ignored because its empirical identification is challenging due to a preponderance of zero-AVC segments.

There is limited literature on modeling time-varying parameters in discrete response models, which are essential to capture the temporal variation in covariate effects due to the unobserved factors (e.g., environmental conditions). To this end, previous studies adopted Markov Switching Models (MSMs) in crash frequency models (Malyshkina et al., 2009; Malyshkina and Mannering, 2010), ordered response models for injury severity (Xiong et al., 2014), and multinomial choice models (Bansal et al., 2020). After the formal introduction of the term "temporal instability" by Mannering (2018), most recent analytical studies in accident research check for its presence using likelihood ratio test (Behnood and Mannering, 2019; Islam et al., 2020; Islam and Mannering, 2020, 2021; Yu et al., 2021). However, to conduct such hypothesis testing, the model is required to be estimated as many times as the number of periods. This approach is not feasible to capture temporal instability caused due to seasonality across months of the year. MSMs also offer a very restrictive specification as they only allow parameters to take as many values as the number of latent states, and going beyond two latent states increases model complexity. Therefore, in the proposed model, we capture seasonality through time-varying parameters. To the best of our knowledge, this is the first study that accounts for seasonal effects along with unknown exposure in hot spot identification using large-scale data.

In sum, the proposed model incorporates exposure (i.e., segment-specific animal crossing) using a nonparametric DP mixture, and the number of segment-level AVCs is assumed to follow a binomial distribution. The probability of AVC occurrence is a logistic function of time-varying parameters and segment-specific characteristics. Special attention is paid to AVC seasonality and the preponderance of zero-crash segments while avoiding computational burdens that often accompany discrete response analysis for such a large data set obtained for Texas (roughly 100,000 reasonably homogeneous [in design attributes, like curvature, grade, number of lanes, speed limit, and median presence] segments, as distinguished in the Texas Department of Transportation's state-maintained network). A Pólya-Gamma augmented Gibbs sampler is derived for computationally tractable estimation of the proposed model.

## ANIMAL-VEHICLE COLLISIONS IN TEXAS

The dataset used here comes from two sources. First, AVC records are from the Crash Records Information System (CRIS) maintained by the Texas Department of Transportation. Second, segment-specific roadway design factors were obtained from the Texas Department of Transportation website. Figures 1 and 2 show the 43,319 AVCs that were reported over the 2010-2016 seven-year period.

Figure 1 shows a small increase in total AVCs in more recent years, perhaps as traffic has risen, with the Texas economy bouncing back from a global recession. More interestingly, the months of October through December demonstrate much higher counts. This seasonal pattern comes largely from the white-tailed deer's rutting or breeding season (Bruinderink and Hazebroek, 2003; Niemi et al., 2017; Sullivan, 2011; TPWD, 2019).

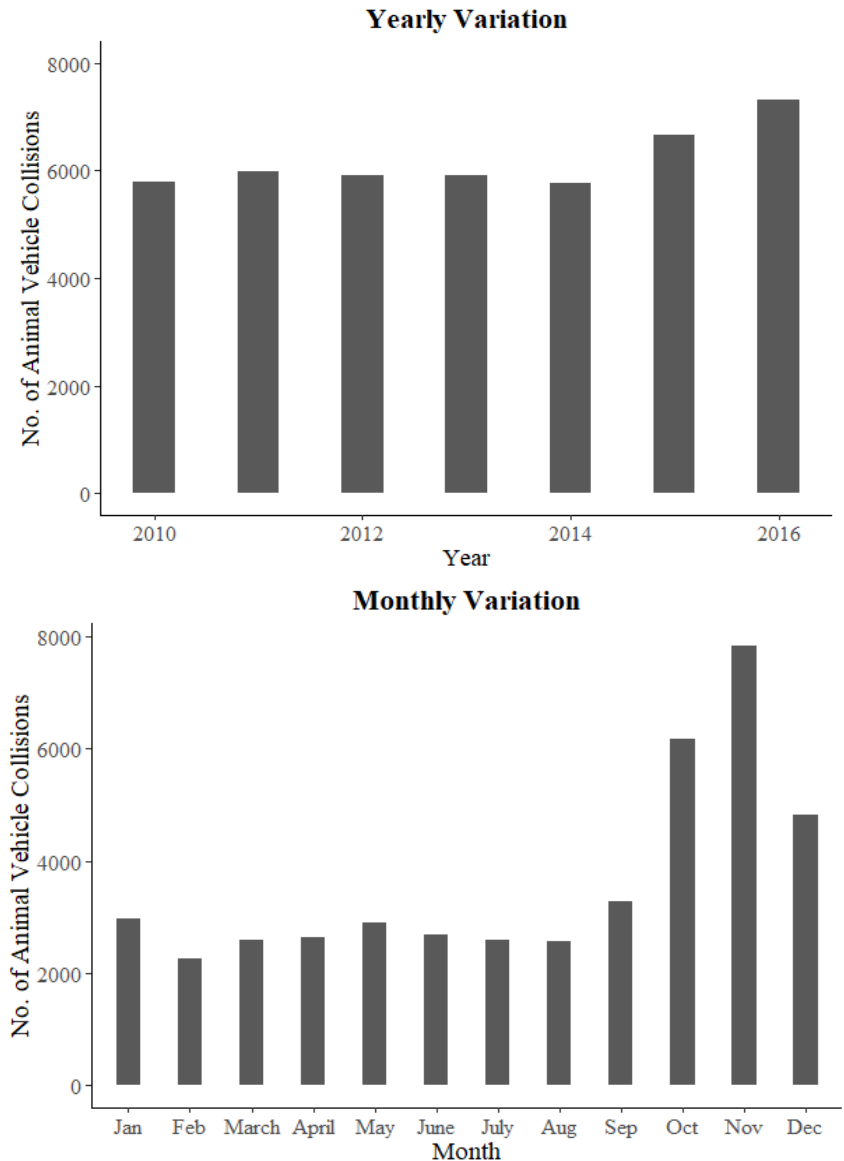

Figure 1: AVC counts by year (left) and month (right), as reported between 2010 and 2016 (on Texas' state-maintained highways)

Figure 2 shows the locations of AVCs associated with the 120,726 segments of state-maintained roadway network[*]. It is clear from the figure that a large portion of the AVCs are located on the east side of the state around several urbanized areas.

Figure 3's upper panel shows that AVCs rarely occur on most segments after disaggregating AVCs over all distinctive Texas highway segments. Further temporal disaggregation to the level of monthly data shows that reported AVC counts are very low along all Texas segments. Figure 3's lower panel shows how just 0.4% (43,324 of 10,140,984) of the monthly segment-level AVC counts are non-zero. Among the non-zero monthly segment-level AVCs (Figure 3's bottom right), only 4.8% (2,085/43,324) have more than one AVC (with a maximum monthly count of 6 AVCs). Accordingly, two important challenges can emerge for segment-level and monthly AVC counts: the computation involved increases dramatically due to the number of observations (120,726 segments × 12 months = 1,448,712), counts are very sparse (typically zero) and highly variable.

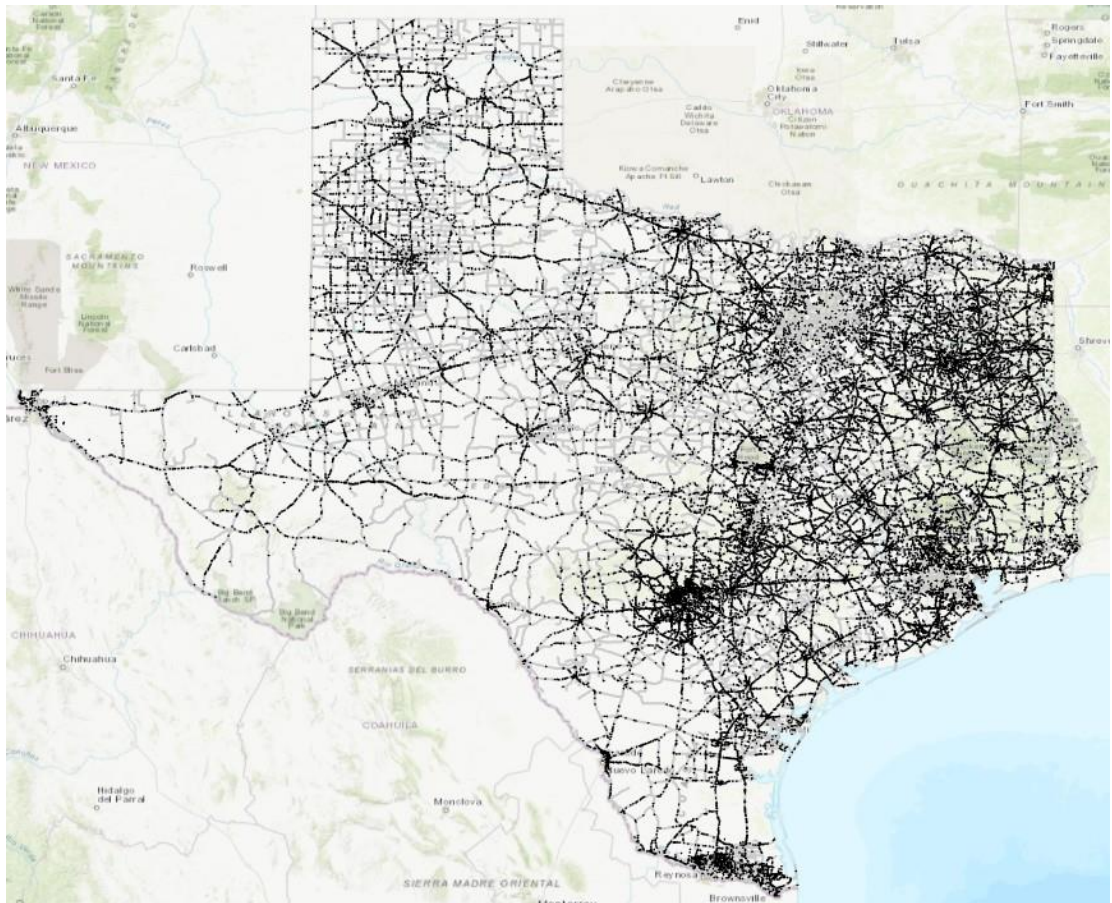

Figure 2: Texas' state-maintained roadway network with reported AVCs (from 2010 through 2016) shown as black dots

---

[*] A small percentage of AVC displayed didn't occur on the network system. The total number of off-system AVCs is 5930, which account for 12 percent of the total AVCs recorded.

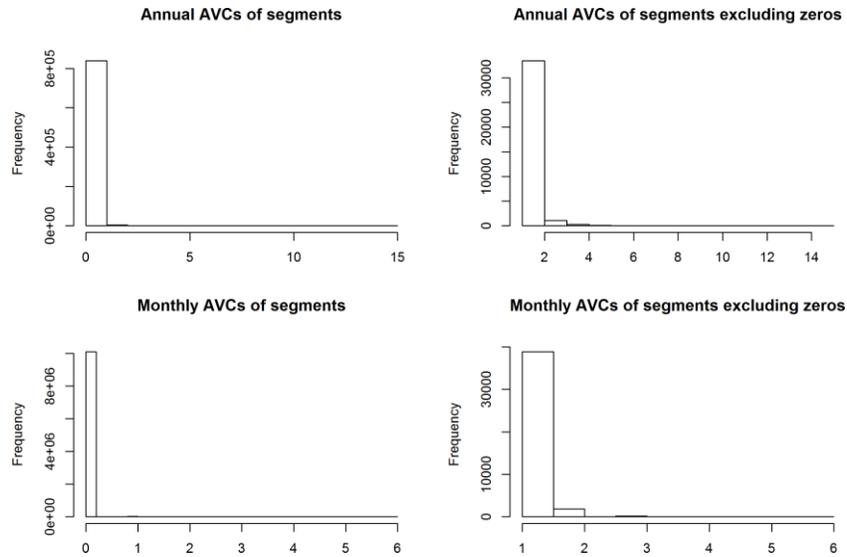
Figure 3: Texas AVC counts by year (top two) and by month (bottom two)

Figure 4 shows AVCs recorded on segments in a small section of Texas. In this figure, the AVCs are denoted as circles (one for each occurrence), whereas segments are shown as solid black lines with their endpoints indicated by crossbars. As evident in Figure 4, most segments have zero AVCs recorded, suggesting that spatial autocorrelation will show as near-zero at this highly disaggregated level although AVC clustering is evident at regional and state levels.

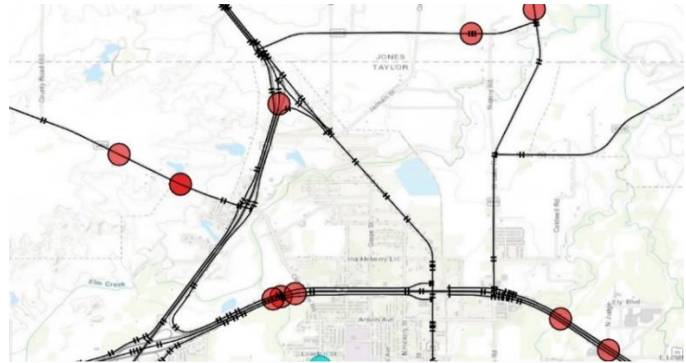
Figure 4: Reported AVCs from 2010 through 2016 in a small area in Texas

Moreover, Figure 4's zero-count segments may indicate heterogeneity across segments. It is possible that many segments located in Figure 4's bottom left (or its top left) are elevated bypasses or have lots of fencing or special underpasses to avoid animals crossing at grade. For this reason, the effect of a specific design factor may only impact AVC counts on the selected subgroup of segments in a large network. In this case, the need to identify the heterogeneity among a large number of segments further complicates the computation.

In summary, the data analysis reveals several important aspects influencing AVC modelling and inference. These include AVC seasonality, the sparse and highly variable nature of AVC count data at the monthly and segment levels, and potential for observed and unobserved heterogeneity among segments. The next section discusses how all these aspects are incorporated in the proposed count data model to find AVC hotspots.

## THE MODELING FRAMEWORK

To account for several important aspects of AVC modelling in a Bayesian hierarchical framework, the model specification begins with the use of a binomial distribution for the number of AVCs recorded at each segment $s$ in month $t$, so that the probability of having $k$ reported AVCs for a segment-time pair $(s, t)$ is

$$P(k_{s,t} \mid n_{s,t}, p_{s,t}) = \binom{n_{s,t}}{k_{s,t}} p_{s,t}^{k_{s,t}} (1 - p_{s,t})^{n_{s,t} - k_{s,t}}, \qquad (1)$$

where $n_{s,t}$ is the number of animal road crossings depending on animal habitats and seasonality, and $p_{s,t}$ is the probability of an AVC occurrence, both of which vary by location ($s$) and month ($t$). Using the binomial distribution with two parameters $n_{s,t}$ and $p_{s,t}$, the number of AVCs, $k_{s,t}$, can be interpreted as the result of repeated $n_{s,t}$ Bernoulli trials. Each trial represents an animal road crossing with probability of $p_{s,t}$ causing an AVC. For this reason, the collision probability, $p_{s,t}$ can be regarded as a quantity that is determined by segment-specific characteristics (like land-use and segment-design factors), and time-varying natural factors (like rainfall) and is assumed to have a logistic functional form:

$$p_{s,t} = \frac{1}{1 + \exp(-\psi_{s,t})}, \qquad (2)$$

where $\psi_{s,t}$ determines the probability of causing an AVC when an animal road crossing is made at segment $s$ and month t, and can be represented as a linear function of characteristics, namely segment-specific design factors like land use, and time-varying natural factors, such as rainfall (see Equation 3). In addition, there may be heterogeneity in collision probability due to the omission of important segment-specific factors. In order to accommodate this possibility, a two-component clustering is incorporated for the intercept:

$$\psi_{s,t} = \alpha_{0,t} I_{s,t} + \boldsymbol{\beta}' \boldsymbol{x}_s + \boldsymbol{\gamma}'_t \boldsymbol{y}_{s,t}, \qquad (3)$$

where $\boldsymbol{x}_s$ is a column vector containing the time-invariant design factors of segment $s$, $\boldsymbol{\beta}$ is the corresponding conformable parameter vector, $\boldsymbol{y}_{s,t}$ is a column vector containing time-varying parameters, and $\boldsymbol{\gamma}_t$ is the corresponding conformable month-specific parameter vector.[†] More importantly, $I_{s,t}$ is an indicator for the non-zero constant effect $\alpha_{0,t}$ at segment $s$ and month t. In other words, the constant effect of each segment-time pair arising from this specification is either zero or $\alpha_{0,t}$. Nonzero $\alpha_{0,t}$ suggests that some network segments have their collision probabilities affected by some important but unknown factors.

While finding appropriate specification of $n_{s,t}$, it is worth noting that the total number of animal crossings for each segment differs across segments and depends on the seasonality of average animal activity levels as seen in Figure 1. To achieve this flexibility, while simultaneously reflecting very low AVC counts on most segments, a nonparametric Dirichlet process (DP) prior is used for the number of animal road crossing at all segments ($s$) and at all months ($t$) of a year: $n_{s,t}$. In summary, the proposed modelling framework for

---

[†] Time-varying parameters on time-invariant attributes can be easily incorporated in the proposed model, but are not specified here to avoid explosion of the parameter space. We could afford time-varying coefficients on time-varying attributes since there is just one attribute (rainfall) with monthly variation in this data set.

AVCs is presented below:

$$\begin{aligned}
P &\sim \mathbf{DP}(\vartheta P_0), & q_t &\sim \mathbf{Beta}(a_0, b_0), \\
(\mu_{s,t}, \sigma_{s,t}) &\sim P, & I_{s,t} &\sim \mathbf{Bernoulli}(q_t), \\
n^*_{s,t} &\sim \mathbf{Normal}_{(-0.5,\infty)}(\mu_{s,t}, \sigma^2_{s,t}), & \alpha_{0,t}, \boldsymbol{\beta}, \boldsymbol{\gamma}_t &\sim \mathbf{MVN}(\mathbf{0}, \boldsymbol{\Sigma}_{0,t}), \\
n_{s,t} &= \lfloor n^*_{s,t} \rfloor, & p_{s,t} &= 1/(1 + exp(-\psi_{s,t})), \\
& & \psi_{s,t} &= \alpha_{0,t} I_{s,t} + \boldsymbol{\beta}' \boldsymbol{x}_s + \boldsymbol{\gamma}'_t \boldsymbol{y}_{s,t},
\end{aligned} \quad (4)$$

$$k_{s,t} \sim \mathbf{Binomial}(n_{s,t}, p_{s,t}).$$

The number of animal road crossings, $n_{s,t}$ and the collision probability, $p_{s,t}$, on segment $s$ in month $t$ are obtained from the left and the right blocks of Equation 4, respectively. The number of observed AVCs on segment $s$ in month $t$, $k_{s,t}$, is a realization of the binomial distribution with the parameters $n_{s,t}$ and $p_{s,t}$, as shown in the last part of Equation 4.

More specifically, in the top-left block of the equations, a discrete distribution $P$ is drawn from DP with scalar precision parameter $\vartheta$ and base distribution $P_0$. Then the cluster locations $\mu_{s,t}$ and scales $\sigma_{s,t}$ are generated from the discrete distribution $P$ for each segment $s$ and month $t$. Conditional on the cluster locations and scales, a real-valued latent quantity $n^*_{s,t}$ is drawn. Then, the total number of animal crossings, $n_{s,t}$, at site $s$ and month $t$ is set equal to the nearest integer, $\lfloor n^*_{s,t} \rfloor$. The truncated normal distribution on $n^*_{s,t}$ ensures the non-negativity of $n_{s,t}$.

For the collision probability, $p_{s,t}$, specification in the top-right block of Equation 4, the indicator probability, $q_t$, is first drawn from a Beta distribution with prior parameters $a_0$ and $b_0$ for each month $t$. Then the indicator variable, $I_{s,t}$, for all segments and months is generated from a Bernoulli distribution using this indicator probability, $q_t$. The non-zero constant effect, the effect of segment-specific design factors, and time-varying natural factors, $[\alpha_{0,t}, \boldsymbol{\beta}', \boldsymbol{\gamma}_t']$ are drawn from a uninformative multivariate normal (MVN) distribution with prior mean zero (**0**) and diagonal covariance, $\boldsymbol{\Sigma}_{0,t}$. Then $\psi_{s,t}$ is determined by the dot product of attributes $[I_{s,t}, \boldsymbol{x}'_s, \boldsymbol{y}_{s,t}']$ and parameters $[\alpha_{0,t}, \boldsymbol{\beta}', \boldsymbol{\gamma}_t']$, which is further transformed to the collision probability, $p_{s,t}$, after passing through a logistic function.

The proposed hierarchical model was estimated using a Markov Chain Monte Carlo simulation. Algorithm 1 shows the step-by-step sampling from the conditional posterior distributions. Key features to note are the Pólya-Gamma data augmentation step to address the non-conjugacy of the logistic probability function (Polson et al., (2013)) and the use of a stick-breaking construction to obtain the DP prior (Canale and Dunson, (2011)). The complete derivation of the Gibbs sampler is provided in the Appendix.

**Initialize parameters** – clusters $\{1, \cdots, C\}$, latent variables, and hyper-parameters

**Step 1: Draw $n_{s,t}$ using a Metropolis-Hastings (MH) step**

→ Step 1a: Assign cluster ID to each segment-month $(s, t)$ pair.

→ Step 1b: Update segment-specific parameters for each time period $(\mu_{s,t}, \sigma_{s,t}^2)$ from multinomial distribution using cluster parameters $\mu_l^*, \sigma_l^{*2}$ as

$$p(\mu_{s,t} = \mu_c^* \text{ and } \sigma_{s,t}^2 = \sigma_c^{*2} | \cdot) = \frac{w_c p(n_{s,t}|\mu_c^*, \sigma_c^{*2})}{\sum_{l=1}^{C} w_l p(n_{s,t}|\mu_l^*, \sigma_l^{*2})},$$

where $p(n_{s,t}|\mu_l^*, \sigma_l^*) = \frac{\Phi(n_{s,t}+1/2|\mu_l^*,\sigma_l^{*2}) - \Phi(n_{s,t}-1/2|\mu_l^*,\sigma_l^{*2})}{1 - \Phi(-1/2|\mu_l^*,\sigma_l^{*2})}$, and $\Phi(.)$ is a normal cumulative distribution function.

→ Step 1c: Update cluster weights $w_l$ using a stick-breaking construction with Beta-distributed $V_l$ where $V_l|. \sim \textbf{Beta}(1 + n_l, \vartheta + \sum_{i=l+1}^{C} n_i)$, $w_1 = V_1, w_l = V_l \prod_{i<l}(1 - V_i)$ for $l = 2, \dots, C$, and $n_l$ is the number of $\mu_{s,t}$ that is equal to $\mu_l^*$. (see [Appendix](#) for more details on $V_l$)

→ Step 1d: Set $n_{s,t}^* = \Phi^{-1}(u_{s,t} | \mu_{s,t}, \sigma_{s,t}^2)$,

where $u_{s,t} \sim \textbf{Uniform}\left(\Phi\left(n_{s,t} - \frac{1}{2}|\mu_{s,t}, \sigma_{s,t}^2\right), \Phi\left(n_{s,t} + \frac{1}{2}|\mu_{s,t}, \sigma_{s,t}^2\right)\right)$

→ Step 1e: Update cluster $(\mu_l^*, \sigma_l^{*2})$ from the Normal-Gamma distribution as

$(\sigma_l^*)^{-2}|. \sim \textbf{Gamma}\left(a_0 + \frac{n_l}{2}, b_0 + \frac{1}{2}\sum_{\{(s,t):\mu_{s,t}=\mu_l^*\}}\left((n_{s,t}^* - \eta) + \frac{n_l}{1+n_l}\eta^2\right)\right)$, and

$\mu_l^*|. \sim I_{\left[-\frac{1}{2}, \infty\right)} \textbf{Normal}\left(\frac{\sum_{\{(s,t):\mu_{s,t}=\mu_l^*\}} n_{s,t}^*}{1+n_l}, \frac{\sigma_l^{*2}}{1+n_l}\right)$.

→ Step 1f: Metropolis-Hastings step with

$$P(n_{s,t}|.) \propto \left[\sum_{l=1}^{C} w_l \frac{\Phi\left(n_{s,t}+\frac{1}{2}|\mu_l^*, \sigma_l^{*2}\right) - \Phi\left(n_{s,t}-\frac{1}{2}|\mu_l^*, \sigma_l^{*2}\right)}{1-\Phi\left(-\frac{1}{2}|\mu_l^*, \sigma_l^{*2}\right)}\right] \times \textbf{Binomial}(k_{s,t}| n_{s,t}, p_{s,t}).$$

**Step 2: Draw $p_{s,t}$**

→ Step 2a: Draw auxiliary variable $\omega_{s,t}|. \sim \textbf{PólyaGamma}(n_{s,t}, \alpha_{0,t}I_{s,t} + \boldsymbol{\beta}'\boldsymbol{x_s} + \boldsymbol{\gamma}_t'\boldsymbol{y_{s,t}})$.

→ Step 2b: Draw $\boldsymbol{\beta}|. \sim \textbf{MVN}(\boldsymbol{m_\beta}, \boldsymbol{V_\beta})$,

where, $\boldsymbol{V_\beta} = \left(\sum_t \sum_s (\omega_{s,t} \boldsymbol{x_s} \boldsymbol{x_s'}) + \boldsymbol{B_0}^{-1}\right)^{-1}$ and
$\boldsymbol{m_\beta} = \boldsymbol{V_\beta}\left(\sum_t \sum_s \boldsymbol{x_s}(\kappa_{s,t} - \omega_{s,t}\boldsymbol{\gamma}_t'\boldsymbol{y_{s,t}} - \omega_{s,t}\alpha_{0,t}I_{s,t})\right)$.

→ Step 2c: Draw $\alpha_{0,t}, \boldsymbol{\gamma_t}|. \sim \textbf{MVN}(\boldsymbol{m_t}, \boldsymbol{V_t})$,

where, $\boldsymbol{V_t} = \left(\sum_s (\omega_{s,t} \boldsymbol{z_{s,t}} \boldsymbol{z_{s,t}'}) + \boldsymbol{D_0}^{-1}\right)^{-1}$ and $\boldsymbol{m_t} = \boldsymbol{V_t}\left(\sum_s \boldsymbol{z_{st}}(\kappa_{s,t} - \omega_{s,t}\boldsymbol{\beta}'\boldsymbol{x_s})\right)$.

→ Step 2d: Draw $I_{s,t}$ with

$P(I_{s,t} = 1|\cdot) = \frac{P_{s,t}^1 q_t}{P_{s,t}^0(1-q_t) + P_{s,t}^1 q_t}$, where

$P_s^0 = \exp\left(\kappa_{s,t}\psi_{s,t} - \frac{1}{2}\omega_{s,t}\psi_{s,t}^2\right) \ | \psi_{s,t} = \boldsymbol{\beta}'\boldsymbol{x_s} + \boldsymbol{\gamma'}_t\boldsymbol{y_{s,t}}$ and

$P_s^1 = \exp\left(\kappa_{s,t}\psi_{s,t} - \frac{1}{2}\omega_{s,t}\psi_{s,t}^2\right) \ | \psi_{s,t} = \alpha_{0,t}I_{s,t} + \boldsymbol{\beta}'\boldsymbol{x_s} + \boldsymbol{\gamma}_t'\boldsymbol{y_{s,t}}$.

→ Step 2e: Draw $q_t|. \sim \textbf{Beta}(1 + \sum_s I_{s,t}, 1 + \sum_s (1 - I_{s,t}))$.

Algorithm 1: Blocked Gibbs Sampler for Detecting Animal-Vehicle Crossings and Collisions
(Note: All draws are for each segment, $s$, and month, $t$, as applicable.)

**RESULTS**

The model described above was estimated using the Texas AVC dataset of the year 2016. Table 1 provides a summary of the segment-level design and environmental features that were used as explanatory variables.

Table 1: Summary statistics for segment-specific factors considered in the model

| Variable Name | Min. | Mean | Max. |
|---|---|---|---|
| Segment Length (miles) | 0.001 | 0.66 | 30.11 |
| Average Daily Traffic, ADT (1000 vehs/day) | 0.01 | 11.20 | 341.30 |
| # Lanes | 1 | 2.77 | 14 |
| Roadbed Width (ft) | 10 | 47.11 | 318 |
| Median Width (ft) | 0 | 7.45 | 710 |
| Inside Shoulder Width (ft) | 0 | 5.49 | 60 |
| Outside Shoulder Width (ft) | 0 | 6.53 | 53 |
| Surface Width (ft) | 10 | 33.82 | 236 |
| % K Factor (for traffic peaking) | 4.20 | 10.75 | 19.90 |
| Controlled Access? | 0 | 0.18 | 1 |
| Posted Speed Limit (mph) | 5 | 56.15 | 85 |
| Urban Area? | 0 | 0.29 | 1 |
| Median Present? | 0 | 0.04 | 1 |

Algorithm 1 was used to draw 20,000 MCMC samples from the conditional posterior distributions of the model parameters, of which first 10,000 burn-in samples were discarded. The estimation took about 24 hours on a high-performance computer with six Intel Xeon cores operating at 3.4 GHz and having 128 GB of RAM. MCMC chains converged with an average Gelman & Rubin R-hat diagnostic of 1.011. Figure 5 shows the posterior expected AVCs compared against observed AVCs by month for the year 2016. The relative magnitude of the aggregated quantities resembles the seasonal pattern in the observed data. Such resemblance illustrates good predictive performance of the proposed model.[‡]

Table 2 shows the estimates and statistics associated with parameters ($\alpha_{0,t}, I_{s,t}$ and $\boldsymbol{\gamma_t}$), which form the time-varying component of the collision probability link function. Three main insights are drawn from these estimates. First, around 53% of the segments have non-zero $I_{s,t}$, i.e. they carry an inherent non-zero constant effect (unexplained by the observed covariates) on the AVC probability. The posterior mean of this non-zero constant effect ($\alpha_{0,t}$) is negative and the magnitude ranges between 0.2 and 0.5 for most months. Second, Table 2 also shows that around 0.8% to 1.5% of segments have non-zero animal crossings ($n_{s,t}$) in each month, which is consistent with small proportion of segments with non-zero AVCs in the data. This result implies the DP process could properly cluster segments with no exposure. Third, rainfall effect on collision probability ($\boldsymbol{\gamma_t}$) is generally positive and is statistically significant in March-April and September-November.

---

[‡] Figure 5's absolute magnitudes are generally smaller than those shown in
Figure 1 since Figure 5 estimates the sums over all segments for a given year, whereas Figure 1 aggregates AVCs across segments and over seven years.

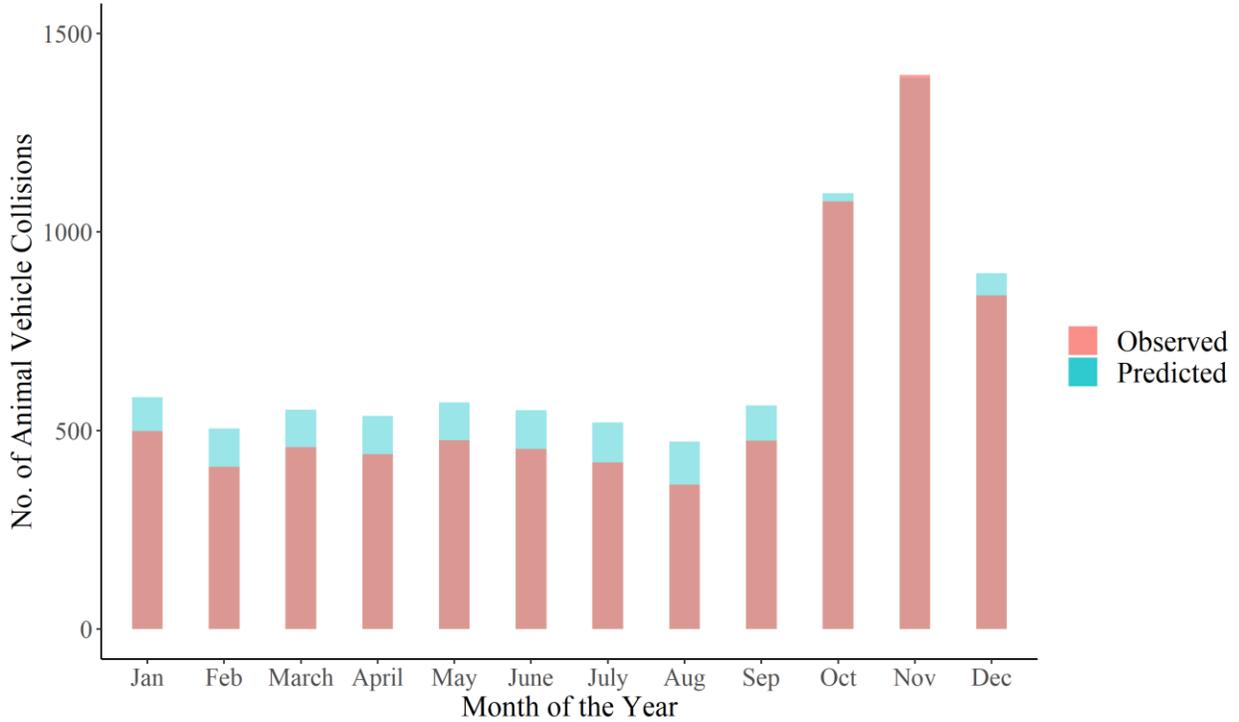

Figure 5: Posterior Expected Versus Observed Number of Animal-Vehicle Collisions by Month.

Table 2: Time-Varying Quantities and Parameter Estimates

| Month | Percent of Non-Zero $n_{s,t}$ | Percent of Non-Zero $I_{s,t}$ | Posterior mean of constant effect ($\alpha_{0,t}$) | Posterior mean of Rainfall (inches) effect ($\gamma_t$) |
|---|---|---|---|---|
| January | 0.839% | 52.9% | -0.296** | 0.113 |
| February | 1.297% | 53.0% | -0.429** | 0.123 |
| March | 1.525% | 52.9% | -0.374** | 0.135* |
| April | 1.115% | 52.7% | -0.392** | 0.214** |
| May | 0.839% | 52.7% | -0.387** | 0.124 |
| June | 1.297% | 53.2% | -0.381** | 0.021 |
| July | 1.525% | 53.0% | -0.431** | -0.028 |
| August | 1.115% | 52.6% | -0.538** | 0.042 |
| September | 0.839% | 52.7% | -0.336** | 0.256** |
| October | 1.297% | 53.6% | 0.266** | 0.328** |
| November | 1.525% | 54.0% | 0.417** | 0.129** |
| December | 1.115% | 54.0% | 0.090 | 0.014 |

** indicates statistical significance, i.e. zero does not lie in 95% credible interval.

Table 3 shows the posterior mean estimates and 95% credible intervals for homogeneous probability parameter ($\beta$). The estimated effects for several design factors are also insightful. Namely, speed limit is positively associated with higher probability of causing an AVC by an animal road crossing. Segments in urban areas tend to have lower probabilities of observing an AVC by animal-road crossings. A median barrier tends to decrease the probability of an AVC, perhaps because animals are unable to see the other side of the segment and may not cross at such locations. Large median widths correspond to a higher likelihood of an AVC, which can correspond to a correlated increase in roadbed width. Outside shoulder widths also show a similar trend, but the inside shoulder width negatively impacts AVC probability. The negative posterior estimate may be due to driving behavioral differences, especially the ability to swerve out of the main lanes and stop at the shoulder leading to lower AVCs, on average. Busy segments or those with a continuous peak traffic flow have a lower likelihood of AVCs, and an increase in average daily traffic corresponds to an increase in AVCs when controlling for several opposing trends (like peak factor and urban areas, for example). Controlled access highways like freeways have a significantly lower likelihood of an AVC, as expected. Land use characteristics also have a significant effect on AVC probability. Segments located near open areas, adjacent to water bodies, or surrounded by trees are more likely to observe an AVC compared to segments near buildings. Interestingly, population density is positively correlated with the likelihood of an AVC while other urban parameters suggest the opposite. An interaction of this variable with urban areas may provide more clarity on the direction of this effect, with population density likely key in co-locating AVCs outside of an urban area.

Table 3: Posterior summary of the homogeneous probability parameters ($\beta$)

| Quantity or Variable | Mean | 95% Credible Interval |
|---|---|---|
| Segment Length (mi) | 37.504 | (34.137, 40.958) |
| Average Daily Traffic (in 1000 vpd) | 0.003 | (3.59E-4, 0.007) |
| Median Width (ft) | 0.009 | (0.005, 0.013) |
| Inside Shoulder Width (ft) | -0.017 | (-0.034, 0.001) |
| Outside Shoulder Width (ft) | 0.076 | (0.058, 0.093) |
| Surface Width (ft) | 0.015 | (0.011, 0.020) |
| Peak Period (%) | -0.075 | (-0.096, -0.053) |
| Controlled Access? | -1.495 | (-1.767, -1.219) |
| Speed Limit (mph) | 0.028 | (0.022, 0.033) |
| Urban Area? | -0.623 | (-0.76, -0.485) |
| Barrier Median Present | -0.455 | (-0.749, -0.166) |
| Terrain Composition: % Water | 0.012 | (0.007, 0.017) |
| % Trees | 0.017 | (0.013, 0.021) |
| % Open Land | 0.008 | (0.005, 0.012) |
| Pop. Density (persons/sq.mi.) | 0.035 | (-0.001, 0.072) |

Using the posterior draws from conditional distribution of probability parameters, the probability of observing an AVC from an animal road crossing on each segment can be evaluated. Figure 6 shows the probability distribution for January and October to observe the difference in crash probabilities by month. The variation in collision probability for both months across segments can be attributed to the differences in road design factors. The likelihood of observing a crash is markedly higher in October than in January, further confirming the proposed model could capture seasonality.

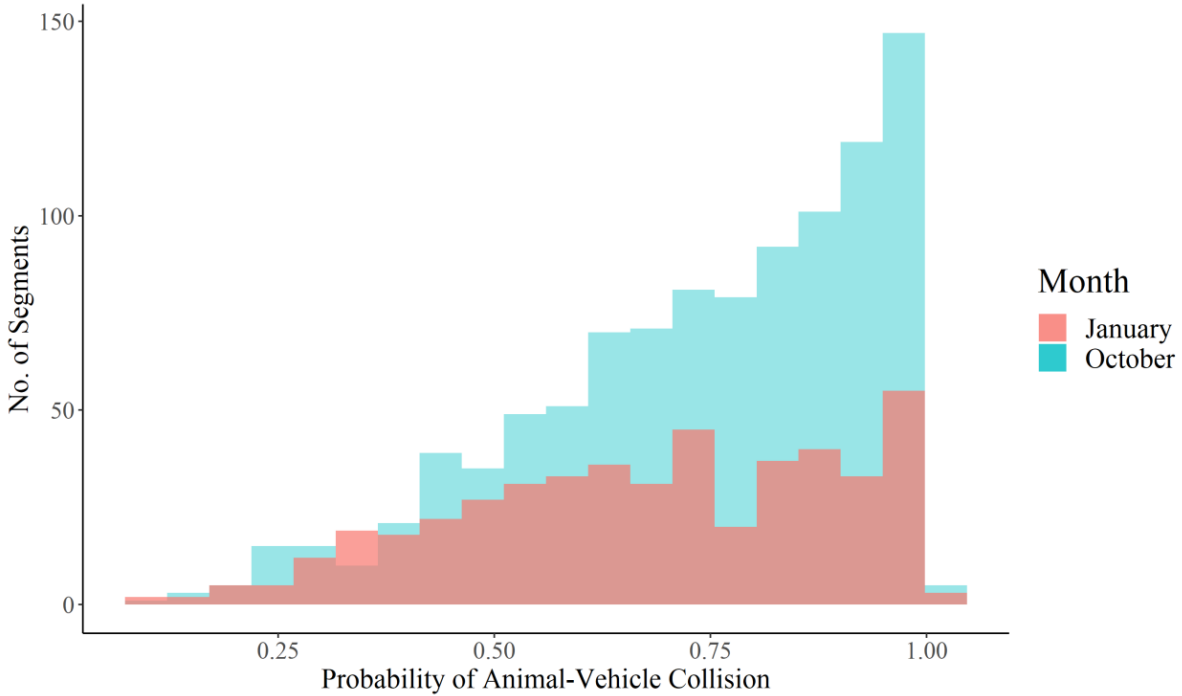

Figure 6: Empirical Posterior Distribution of Collision Probability, $p_{s,t}$ for January and October

In expectation, the occurrence of AVC should be relatively rare, meaning that the probability should be much smaller. However, Figure 6 also indicates that a large proportion of segments is estimated to have the probability above 0.75, meaning that observing an AVC is high conditional on animal road crossings. Such high probability values may seem counter intuitive. In order to make sense of the probabilities shown in Figure 6, it is important to recognize the difference between the information contained in the AVC dataset and that used in forming our expectation. The animal crossing inferred by the model are in fact those crossings that occurred when there are vehicles driving on the segment because only those events will qualify to be "trials" in the Binomial model. There may be other animal road crossings occurring when there are no vehicles on the segments, and these would not lead to AVCs. Therefore, they will not be recorded in the dataset and the exposure model will not be able to detect them. In this sense, the probabilities shown in Figure 6 is in fact the probability of observing an AVC given both the presence of an animal road crossing and vehicles driving on the segment.

The spatial distribution of posterior collision probabilities ($p_{s,t}$) is shown in Figure 7 across the Texas highway network. Figure 7 shows several clusters of light-colored segments, which correspond to the network around major urban areas. In contrast, darker segments are typically major highways that span the entire state. This pattern is a manifestation of the sampled parameters shown in Table 3 (i.e., the negative estimates of urban area, and positive estimates of speed limit and highway type).

The probabilities shown in Figure 7 represent the likelihood of causing an AVC by an animal road crossing. This information is useful for appreciating AVC contributions of road design decisions. However, the number of road crossings by animals on any given segment is also key in determining expected AVCs. Segments with high crash probability, but having zero or very few animal road crossings are relatively less of a concern as compared to segments having a high crash probability and a large number of animal crossings. The segments of the latter type can be regarded as hotspots, meriting AVC prevention considerations. Identifying such hotspots can guide investments and intervention decisions. The posterior mean of the expected AVCs ($n_{s,t} \times p_{s,t}$) are relevant, and are shown in Figure 8 for two different months of a year.

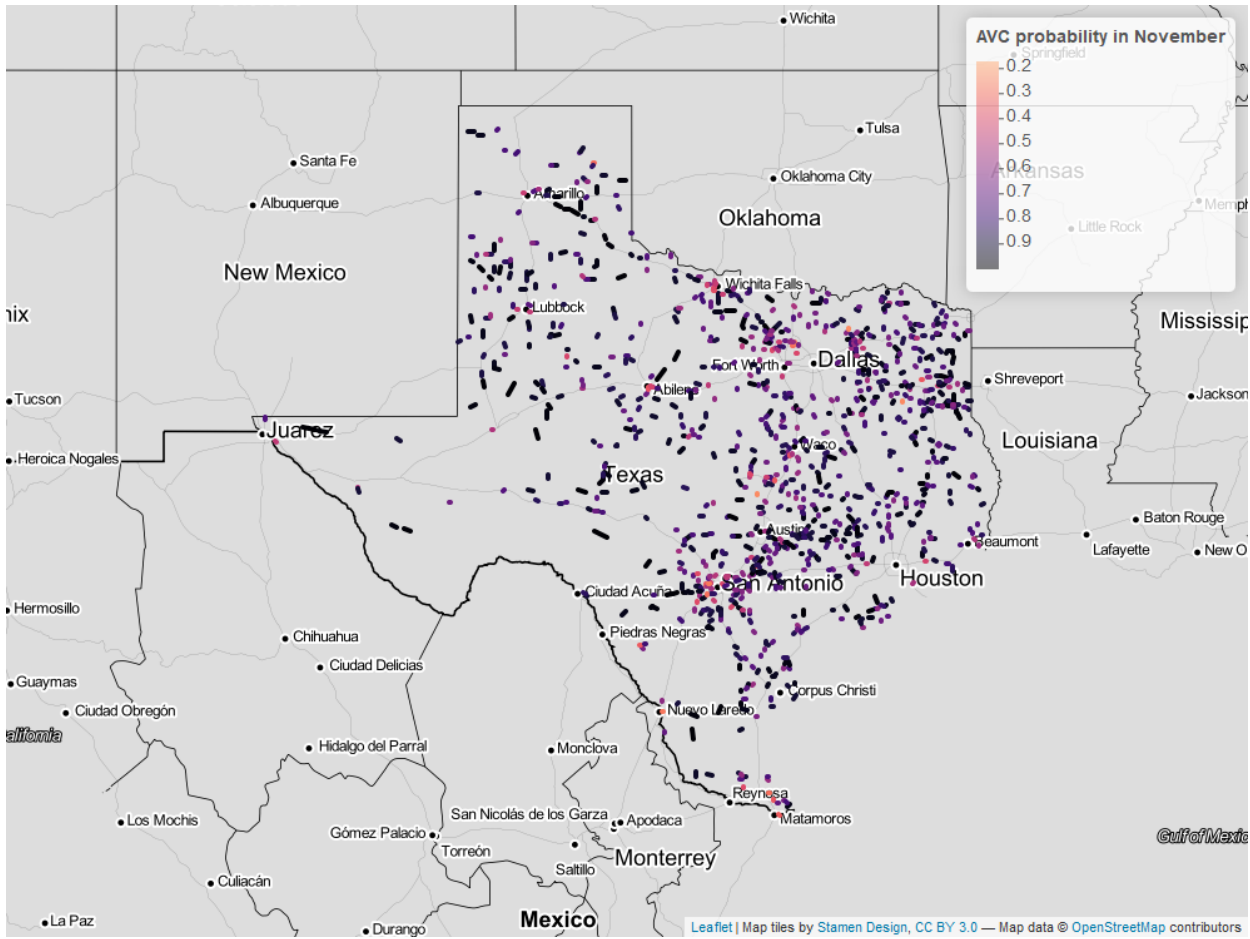

Figure 7: Posterior Mean of Collision Probability $p_{s,t}$ for All Segments in the Texas Roadway Network for November

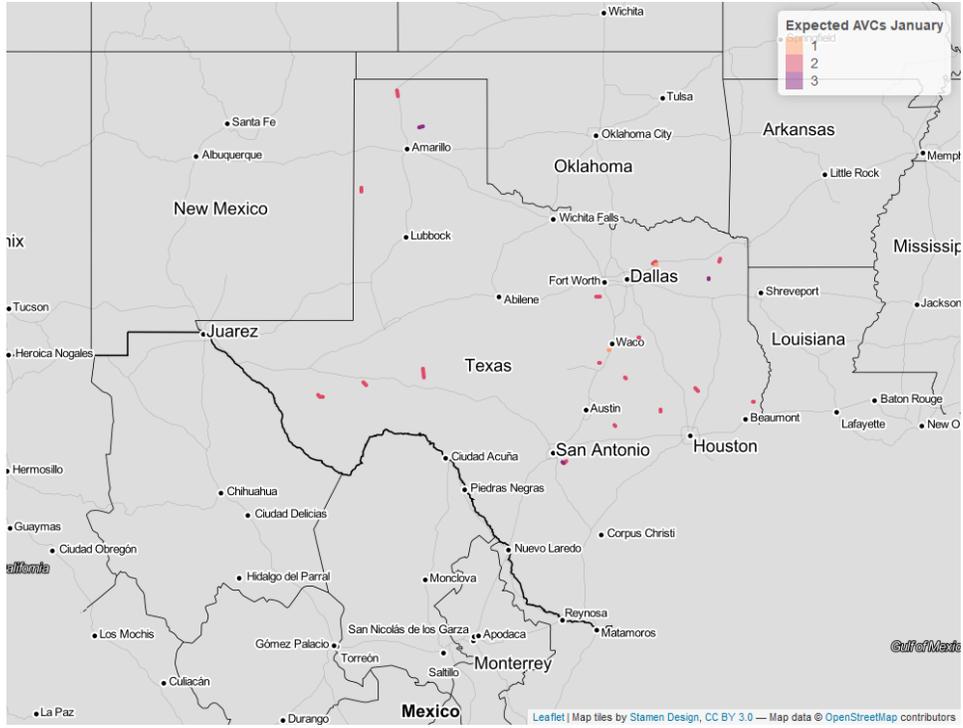

(a) January

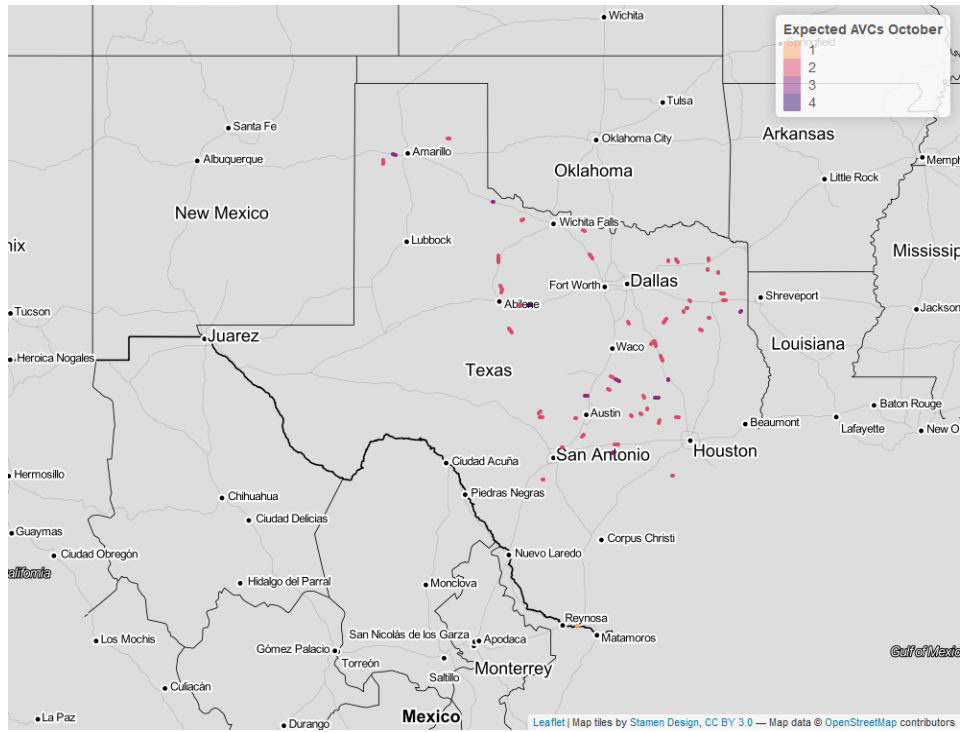

(b) October

Figure 8: Posterior Expected AVCs for All Segments in the State-Controlled Texas Roadway Network

By using the number of crossing ($n_{s,t}$) in the calculation for the expected AVCs, Figure 8 shows the prominent effect of seasonality. More specifically, the expected AVC values are higher (darker colored) for more segments in October than that in January. This corresponds well with

Figure 1's patterns and allows the model-identified hotspots to vary by month or season. Another apparent feature evident in Figure 8 is that segments with higher expected AVCs are only a small share of all segments, and scattered across the Texas network. This small share is due to the fact that over 98% of $n_{s,t}$ values are zero in any month (as noted in Table 2).

Since segments with a high number of expected AVCs are scattered across the network, a smooth change in expected AVCs from one segment to other nearby segments may be preferred, through spatial autocorrelation. However, in doing so the advantage of having segments with zero animal crossings play no role in the parameter estimation is lost, causing parameter estimates to be biased low from averaging effects over nearby segments. Moreover, this contradicts the goal of inferring segment-specific design factor effects and shifts the focus onto higher levels of spatial aggregation.

Given a segment-level focus of this analysis, another interesting aspect that is easy to identify from this model is the effect of a change in any of the design factors. Among Table 3's design factors, speed limit is the most cost-effective (and implementation-time-efficient) way to lower AVCs. The posterior means of collision probabilities are recalculated when assuming a 10 mph decrease in posted speed limit. Figure 9 shows the histogram of change in collision probability resulting from this new speed limit. It is interesting to see that there is a stark difference in the effect of lowering speed limits in different months. This shows the temporal effectiveness of such a countermeasure. Other design factors can be tested for a more thorough temporal benefit, and the State of Texas may wish to focus on those, to save human and animal lives, while avoiding injuries and property loss. Figure 10 shows the spatial nature of the decrease in expected AVCs ($n_{s,t} \times p_{s,t}$) where marginally larger reductions in expected AVCs are observed in urban areas.

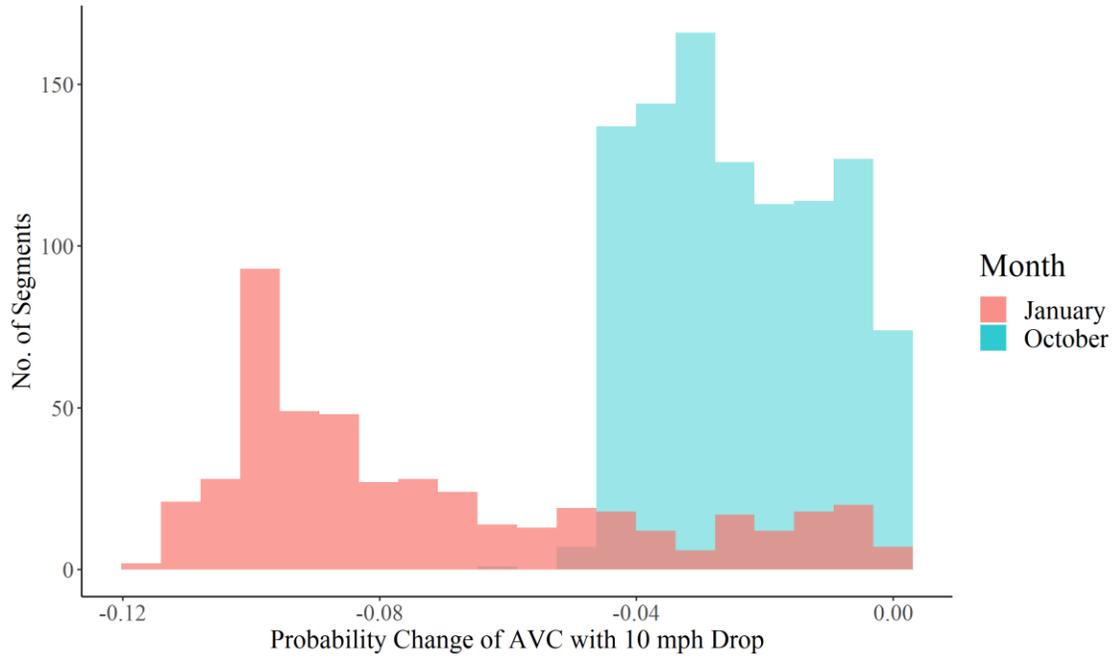

Figure 9: Change in Posterior Probability of AVC for January and October with a 10 mph Speed Limit Drop Across Texas Highways

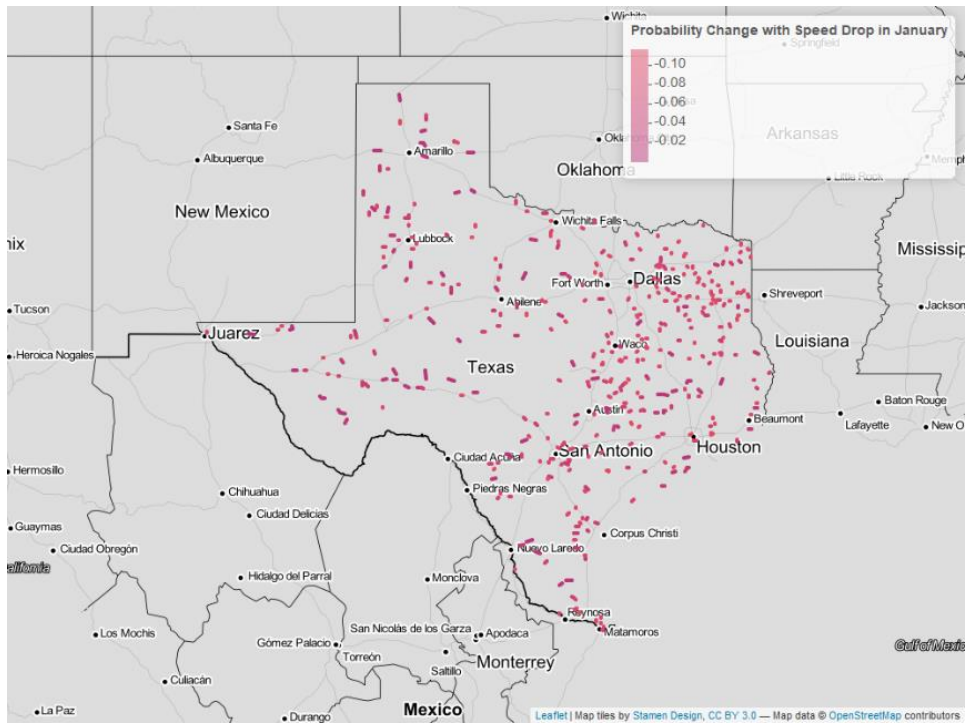

Figure 10: Posterior Mean of Changes for Expected AVCs following a 10 mph Speed Limit Reduction on Texas Highways

**CONCLUSIONS**

This study develops and demonstrates a new method for analysis of low count values across many points in time and space, using a large network of roadway segments. The model is validated using Texas' animal-vehicle collision (AVC) dataset. The proposed model uses binomial distribution to specify the AVC counts and allows the number of animal crossings to be governed by a Dirichlet process (DP). The collision probability is represented using a logistic function that depends upon segment-specific factors, monthly rainfall, and segment-month random effects. DP enables the modelling of segments with zero AVCs because it creates clusters of segments nonparametrically that can share information. Time-varying probability specification helps in capturing seasonal effects. To address the non-conjugacy of posterior updates of the parameters associated with the logistic probability function, Pólya-Gamma data augmentation is adopted.

Several advantages of the proposed modeling framework become clear in the case study. First, this new specification enables hotspot identification over time points (for example, months or seasons), not just space. Second, the impacts of various segment-specific attributes are inferred directly across all 120,726 locations. The proposed modeling framework thus allows policymakers to dive deep into factors that impact AVCs. The impact of purposeful modifications in any segment-specific factor (like speed limit or lane width) on AVC counts can be estimated in a relatively straightforward fashion. However, to make the marginal effect estimates for control variables (e.g., speed limits or lane width) meaningful to devise AVC avoidance policies, various other traffic-related measures need to be considered.

In summary, AVCs are challenging to predict due to the interactions of complex vegetative, climatic, traffic, and human factors. The inclusion of more variables like driver sight distances, use of fencing, availability of underground tunnels for animal crossings, and clear zone dimensions alongside highways may be helpful. Extending the model to include the time of day variability can also help improve estimates since a large proportion of AVCs occurs at night[§]. Moreover, accounting for unobserved heterogeneity in the effect of covariates on crash probability is also likely to improve prediction accuracy. However, increased model complexity will require advanced techniques to speed up model estimation and convergence, such as the use of Variational Bayes (Bansal et al., 2021).

---

[§] FHWA study (https://www.fhwa.dot.gov/publications/research/safety/humanfac/94156.cfm) suggests a large proportion of AVCs occur early in the day between 4 and 6 am and at night between 6 and 11 pm.

**APPENDIX: DERIVATION OF THE GIBBS SAMPLER**

The sampling from conditional posterior distributions in the MCMC estimation of the proposed hierarchical model can be divided into two blocks. Whereas the first block contains the sampling of the number of animal road crossings, $n_{s,t}$, the second block includes the sampling of collision probability, $p_{s,t}$, and related parameters for all segment-month pairs.

In the first block, a stick-breaking construction is considered for the Dirichlet process that enables the estimation of the probability of each cluster containing segments. These probabilities help estimate cluster parameters, and eventually the continuous $n_{s,t}^*$ that is then truncated and discretized to obtain segment-month-level animal road crossing count $n_{s,t}$. Conditional posterior distributions of all parameters in block 1 are in closed-form, except the final step for which the Metropolis-Hastings algorithm is used. In the second block, since the binomial distribution with logistic probability function does not have a conjugate prior, Pólya-Gamma data augmentation is adopted to transform the model likelihood to the Gaussian likelihood (Polson et al., (2013). For the notational simplicity, $P(A|\cdot)$ is used to denote the probability of $A$ conditioning on the rest of the parameters and data.

*A.1 Posterior sampling of $n_{s,t}$ and the related parameters*

Conditioning on the starting value of exposure, $n_{s,t}$, a blocked Gibbs sampler for the Dirichlet process is used to sample $n_{s,t}$ (Ishwaran and James, 2001a).[5] Here, the kernel function used for representing clusters is a truncated normal density function with the truncation made at -0.5 from below to ensure the non-negativity of resulting $n_{s,t}$. For simplicity in cluster-specific distribution, the precision parameter $\vartheta$ is set to one, with the following base distribution:

$$P_0(\mu, \sigma^2) = \mathbf{Normal}(\mu|\mu_0, \sigma^2)\mathbf{Gamma}(1/\sigma^2\ |d_0, e_0), \tag{5}$$

The prior parameters are chosen to be weakly informative ($\mu_0 = 0$, $d_0 = 2$ and $e_0 = 10$). Due to the discrete nature and the limited number of distinct values of $n_{s,t}$, the maximum number of distinct clusters ($C$) is set to 3.[6] For a thorough review of the stick-breaking construction, readers can refer to Ishwaran and James (2001b). Using the above prior specification, the Gibbs sampler proceeds via the following sampling steps:

- For each $s = 1, ..., S$ and $t = 1, ..., T$, update $\mu_{s,t}$ and $\sigma_{s,t}^2$ by sampling from a multinomial distribution with

$$p\left(\mu_{s,t} = \mu_c^* \text{ and } \sigma_{s,t}^2 = \sigma_c^{*2}\middle|\cdot\right) = \frac{w_c p(n_{s,t}|\mu_c^*, \sigma_c^{*2})}{\sum_{l=1}^{C} w_l p(n_{s,t}|\mu_l^*, \sigma_l^{*2})}, \tag{6}$$

  where $w_l$ is the weight and $\mu_l^*$ and $\sigma_l^*$ are parameters for cluster $l$. The kernel function for each cluster is as follows:

---

[5] See Shirazi et al. (2016) for an application in safety research.
[6] We tried to estimate the model with the higher number of clusters, but convergence issues are encountered due to a limited range of $n_{s,t}$.

$$p(n_{s,t}|\mu_l^*, \sigma_l^{*2}) = \frac{\Phi(n_{s,t} + 1/2|\mu_l^*, \sigma_l^{*2}) - \Phi(n_{s,t} - 1/2|\mu_l^*, \sigma_l^{*2})}{1 - \Phi(-1/2|\mu_l^*, \sigma_l^{*2})}, \quad (7)$$

where $\Phi(.)$ is a normal cumulative distribution function

- A stick-breaking construction of Dirichlet process is used to compute the probabilities or weights for each cluster. Assuming $V_l$ for each cluster $l$ is independent **Beta**$(1, \vartheta)$, the weights are $w_1 = V_1$ and $w_l = V_l \prod_{i<l}(1 - V_i)$ for $l = 2, \ldots, C$. By setting $V_C$ to 1, the weights across all clusters are guaranteed to sum to 1. The posterior of $V_l$ accounts for the number of segment-time pairs that belong to cluster $l$. It can be sampled from:

$$V_l \sim \textbf{Beta}\left(1 + n_l, \vartheta + \sum_{i=l+1}^{C} n_i\right), \text{ for } l = 1, \ldots C - 1, \quad (8)$$

where $n_l$ is the number of $\mu_{s,t}$ that is equal to $\mu_l^*$.

- For each $s = 1, \ldots, S$ and $t = 1, \ldots, T$, draw $n_{s,t}^*$ by first sampling from the following:

$$u_{st} \sim \textbf{Uniform}\left(\Phi\left(n_{s,t} - \frac{1}{2}\Big|\mu_{s,t}, \sigma_{s,t}^2\right), \Phi\left(n_{s,t} + \frac{1}{2}\Big|\mu_{s,t}, \sigma_{s,t}^2\right)\right), \quad (9)$$

and then set $n_{s,t}^* = \Phi^{-1}(u_{s,t}|\mu_{s,t}, \sigma_{s,t}^2)$.

- Update the cluster-specific parameters using their conditional distributions for $l = 1, \ldots C$:

$$1/\sigma_l^{*2} \sim \textbf{Gamma}\left(a_0 + \frac{n_l}{2}, b_0 + \frac{1}{2}\sum_{\{(s,t):\mu_{s,t}=\mu_l^*\}}\left((n_{s,t}^* - \eta) + \frac{n_l}{1+n_l}\eta^2\right)\right), \quad (10)$$

$$\mu_l^* \sim I_{\left[-\frac{1}{2},\infty\right)}\textbf{Normal}\left(\frac{\sum_{\{(s,t):\mu_{s,t}=\mu_l^*\}} n_{s,t}^*}{1 + n_l}, \frac{\sigma_l^{*2}}{1 + n_l}\right), \quad (11)$$

where $\eta = \sum_{\{(s,t):\mu_{s,t}=\mu_l^*\}} \frac{n_{s,t}^*}{n_l}$.

- The above steps include update of all parameters associated with the distribution of $n_{s,t}$. Conditional on these parameters and the parameters related to collision probability $p_{s,t}$, the conditional marginal probability of $n_{s,t}$ is as follows:

$$P(n_{s,t}|\cdot) \propto \left[\sum_{l=1,\ldots,C} w_l \frac{\Phi\left(n_{s,t} + \frac{1}{2}|\mu_l^*, \sigma_l^{*2}\right) - \Phi\left(n_{s,t} - \frac{1}{2}|\mu_l^*, \sigma_l^{*2}\right)}{1 - \Phi\left(-\frac{1}{2}|\mu_l^*, \sigma_l^{*2}\right)}\right] \textbf{Binomial}(k_{s,t}|n_{s,t}, p_{s,t}) \quad (12)$$

Utilizing the above expression, a Metropolis-Hastings algorithm is used to sample $n_{s,t}$ for all segment-month pairs.

*A.2 Posterior sampling of $p_{s,t}$ and the related parameters*

The posterior sampling for the parameters related to collision probability, $p_{s,t}$, can be divided into two parts: The first part is concerned with the parameters for defining $p_{s,t}$, which are the regression parameters $\alpha_{0,t}, \boldsymbol{\beta}$ **and** $\boldsymbol{\gamma_t}$, and the second part is related to the parameters associated with indicators, $I_{s,t}$.

Conditional on both $n_{s,t}$ and $I_{s,t}$, marginal posterior distributions of $\alpha_{0,t}, \boldsymbol{\beta}$ **and** $\boldsymbol{\gamma_t}$ are not of well-known form. To address this non-conjugacy challenge, a Pólya-Gamma-distributed auxiliary variable, $\omega_{s,t}$, is introduced for each segment-month pair (Polson et al., (2013).[7] After conditioning on $\omega_{s,t}$ and other parameters, the resulting marginal posterior distributions of $\alpha_{0,t}, \boldsymbol{\beta}$ and $\boldsymbol{\gamma_t}$ become Gaussian. The detailed sampling steps are as follows:

- Conditional posterior distribution of $\omega_{s,t}$ is:

$$\omega_{s,t}|. \sim \textbf{PólyaGamma}(n_{s,t}, \alpha_{0,t}I_{s,t} + \boldsymbol{\beta}'\boldsymbol{x}_s + \boldsymbol{\gamma}'_t\boldsymbol{y}_{s,t}), \tag{13}$$

- It is worth nothing that posterior updates for time-invariant, $\boldsymbol{\beta}$, and time-varying parameters, $[\alpha_{0,t}, \boldsymbol{\gamma_t}]$, differ substantially, and therefore, we update them separately as detailed below:

  a. We follow Polson et al. ((2013) to obtain the posterior update for time-invariant parameters, $\boldsymbol{\beta}$, which turns out to be Gaussian:

  $$\boldsymbol{\beta} \sim \textbf{MVN}(\boldsymbol{m}_\beta, \boldsymbol{V}_\beta), \tag{14}$$

  where,

  $$\boldsymbol{V}_\beta = \left(\sum_t \sum_s (\omega_{s,t}\boldsymbol{x}_s\boldsymbol{x}'_s) + \boldsymbol{B}_0^{-1}\right)^{-1},$$

  $$\boldsymbol{m}_\beta = \boldsymbol{V}_\beta \left(\sum_t \sum_s \boldsymbol{x}_s(\kappa_{s,t} - \omega_{s,t}\boldsymbol{\gamma}'_t\boldsymbol{y}_{s,t} - \omega_{s,t}\alpha_{0,t}I_{s,t})\right), \tag{15}$$

  $\kappa_{s,t} = k_{s,t} - \frac{n_{s,t}}{2}$ and $\boldsymbol{B}_0$ is the prior uninformative covariance matrix (subset of $\boldsymbol{\Sigma_{0,t}}$, a diagonal matrix with large values) for $\boldsymbol{\beta}$.

  b. Similarly, time-varying parameters, $[\alpha_{0,t}, \boldsymbol{\gamma}'_t]$, are drawn from:

  $$[\alpha_{0,t}, \boldsymbol{\gamma}'_t] \sim \textbf{MVN}(\boldsymbol{m}_t, \boldsymbol{V}_t) \tag{16}$$

  where,

  $$\boldsymbol{V}_t = \left(\sum_s (\omega_{s,t}\boldsymbol{z}_{s,t}\boldsymbol{z}'_{s,t}) + \boldsymbol{D}_0^{-1}\right)^{-1},$$

  $$\boldsymbol{m}_t = \boldsymbol{V}_t \left(\sum_s \boldsymbol{z}_{st}(\kappa_{s,t} - \omega_{s,t}\boldsymbol{\beta}'\boldsymbol{x}_s)\right), \tag{17}$$

  where $\boldsymbol{z_{st}} = [I_{s,t}, \boldsymbol{y}'_{s,t}]'$ and $D_0$ is the prior uninformative covariance matrix (subset of $\boldsymbol{\Sigma_{0,t}}$, a diagonal matrix with large values) for $[\alpha_{0,t}, \boldsymbol{\gamma_t}]$.

---

[7] See Buddhavarapu et al. (2016) and Buddhavarapu et al. (2020) for applications in safety research.

For all segment-month pairs, the indicator $I_{s,t}$ is drawn from its conditional posterior distribution:

$$P(I_{s,t} = 1|\cdot) = \frac{P^1_{s,t} q_t}{P^0_{s,t}(1 - q_t) + P^1_{s,t} q_t}, \tag{18}$$

where

$$\begin{aligned} P^0_{s,t} &= \exp\left(\kappa_{s,t}\psi_{s,t} - \frac{1}{2}\omega_{s,t}\psi^2_{s,t}\right) \mid \psi_{s,t} = \boldsymbol{\beta}'\boldsymbol{x}_s + \boldsymbol{\gamma}'_t \boldsymbol{y}_{s,t} \\ P^1_s &= \exp\left(\kappa_{s,t}\psi_{s,t} - \frac{1}{2}\omega_{s,t}\psi^2_{s,t}\right) \mid \psi_{s,t} = \alpha_{0,t} I_{s,t} + \boldsymbol{\beta}'\boldsymbol{x}_s + \boldsymbol{\gamma}'_t \boldsymbol{y}_{s,t} \end{aligned} \tag{19}$$

Lastly, assuming prior distribution for $q_t$ to be **Beta**(1,1), its conditional posterior distribution is:

$$q_t \sim \textbf{Beta}\left(1 + \sum_{s=1}^{S} I_{s,t}, 1 + \sum_{s=1}^{S}(1 - I_{s,t})\right). \tag{20}$$